\documentclass{svjour2}

\usepackage{euscript,amssymb,amsmath,graphicx,epsfig}

\newcommand{\be}[1]{\begin{equation}\label{#1}}
\newcommand{\ee}{\end{equation}}
\newcommand{\ba}[1]{\begin{eqnarray}\label{#1}}
\newcommand{\ea}{\end{eqnarray}}
\newcommand{\rf}[1]{(\ref{#1})}
\newcommand{\nn}{\nonumber}

\begin{document}

\title{Inflation due to quantum potential}

\author{Maxim V. Eingorn \and Vitaliy D. Rusov}

\institute{M. Eingorn \at Physics Department, North Carolina Central University,\\ Fayetteville st. 1801, Durham, North Carolina 27707, U.S.A.\\ \\
V. Rusov \at Department of Theoretical and Experimental Nuclear Physics,\\ Odessa National Polytechnic University,\\
Shevchenko av. 1, Odessa 65044, Ukraine\\ \\
\email{maxim.eingorn@gmail.com}\\
\email{siiis@te.net.ua}\\}

\date{Received: date / Accepted: date}

\maketitle

\begin{abstract}
In the framework of a cosmological model of the Universe filled with a nonrelativistic particle soup, we easily reproduce inflation due to the quantum
potential. The lightest particles in the soup serve as a driving force of this simple, natural and promising mechanism. It is explicitly demonstrated that the
appropriate choice of their mass and fraction leads to reasonable numbers of e-folds. Thus, the direct introduction of the quantum potential into cosmology of
the earliest Universe gives ample opportunities of successful reconsideration of the modern inflationary theory.

\keywords{quantum potential \and inflation}
\end{abstract}

%\pacs{98.80.Cq, 95.35.+d%, 14.80.Bn, 14.80.Hv, 14.80.Va
%}

%98.80.Cq Particle-theory and field-theory models of the early Universe (including cosmic pancakes, cosmic strings, chaotic phenomena, inflationary universe, etc.)
%95.35.+d Dark matter (stellar, interstellar, galactic, and cosmological)
%14.80.Bn Standard-model Higgs bosons
%14.80.Hv Magnetic monopoles
%14.80.Va Axions and other Nambu-Goldstone bosons (Majorons, familons, etc.)

%\vspace{0.3cm}

\section*{Introduction}

In the de Broglie-Bohm causal interpretation of quantum mechanics \cite{Holland,Durr,Wyatt} a special role is attributed to the so-called quantum potential
(commonly denoted by $Q$). It was considered for a long time that nontrivial physical features of this quantity represent exclusively a prerogative of quantum
mechanics and have no classical (non-quantum) analogs. However, in the recent paper \cite{Rusov1} (see also \cite{Rusov2} for generalization to the
relativistic case) it was demonstrated that the quantum potential (the quantum dissipation energy) explains successfully the heart of the quantization problem
in classical mechanics.

As an example of applying the quantum potential approach one can cite deriving the quantum hydrodynamics equations and reproducing easily and straightforwardly
the Bogolyubov spectrum of elementary excitations for the Bose-Einstein condensate of an imperfect fluid with pairwise interaction between the particles
\cite{Tsubota,FOOP}.

The other striking example concerns axions (or axion-like particles) being popular dark matter candidates as well as a valid cause of the Sun luminosity and
total solar irradiance variations (see, e.g., \cite{RusovPRD}). The quantum potential is often used for describing the structure formation in the Universe with
their participation \cite{SikiviePRL9,SikiviePRL12}.

Let us raise the following natural question: what can be the cosmological role of the quantum dissipation energy $Q$? In this paper we give the unambiguous
answer: it can be responsible for inflation! We show that the Einstein equations with the corresponding energy-momentum tensor lead to the scale factor being
characterized by the inflationary behavior at the earliest stage of the Universe evolution. Under certain specified conditions this behavior agrees with the
observational requirements. In the light of the Planck results \cite{PLANCK} the proposed scenario can be a deserving attention alternative to the modern
inflationary theories and a possible way out of their difficulties.

%the monopole, horizon, smoothness, and entropy problems

The paper is organized in the following way. In the first section we construct a cosmological model with the quantum potential, derive the Hubble parameter and
demonstrate the inflation possibility in principle. In the second section duration of inflation, the number of e-folds and other important physical quantities
are estimated and restricted. We conclude by collecting the main results in Summary.

%\vspace{0.3cm}

\section*{Cosmological model with quantum potential}

In the beginning let us confine ourselves to the simple case when the earliest Universe is supposed to be filled with only one component, namely, the entirely
nonrelativistic collisionless gas of point-like particles of the equal mass $m$. Then the energy-momentum tensor components have the following form:
\be{1} T^{ik}=(\varepsilon_c+\varepsilon_q+p_q)u^iu^k-p_qg^{ik}\, ,\ee
where $\varepsilon_c$ is the classical (i.e., non-quantum) energy density (the corresponding classical pressure $p_c$ is equal to zero), while $\varepsilon_q$
and $p_q$ represent quantum admixtures. It is known (see, e.g., \cite{Landau,Gorbunov,Rubakov}) that for a pressureless fluid
\be{2} \varepsilon_c(a)=\varepsilon_{0}\left(\frac{a_0}{a}\right)^3=\rho_{\mathrm{ph}}c^2\, ,\ee
where $a(t)$ is the scale factor entering the standard FLRW metric (the spatial curvature is assumed to equal zero for simplicity)
\be{3} ds^2=c^2dt^2-a^2(t)\left(dx^2+dy^2+dz^2\right)\, ,\ee
and $\varepsilon_{0}$ is the value of $\varepsilon_c$ when the value of $a$ is $a_0$. Further, the physical rest mass density of the considered gas
$\rho_{\mathrm{ph}}=Nm/V_{\mathrm{ph}}\sim1/a^3$, where the physical volume $V_{\mathrm{ph}}\sim a^3$ contains $N$ particles on the average. At the same
time, with the help of the known expression for the one-particle quantum potential (see, e.g., \cite{Holland,Durr,Wyatt,Rusov1} and \cite{Rusov2} for
generalization to the relativistic case)
\be{4} Q_1=\frac{\hbar^2}{2m}\frac{1}{\sqrt{\varepsilon_c}}\,\square\left(\sqrt{\varepsilon_c}\right)\, ,\ee
where $\square$ is the D'Alembert operator ($\square f=f_{;i;k}g^{ik}$ for an arbitrary function $f(t,x,y,z)$, semicolons denote covariant derivatives), it is
easy to obtain
\ba{5} \varepsilon_q=\frac{NQ_1}{V_{\mathrm{ph}}}=\frac{\hbar^2}{2m^2c^2}\sqrt{\varepsilon_{c}}\,\square\left(\sqrt{\varepsilon_{c}}\right)=
-\frac{3}{4}\tau^2\varepsilon_{0}\left(\frac{a_0}{a}\right)^3\left(\frac{\ddot a}{a}+\frac{1}{2}\frac{\dot a^2}{a^2}\right)\, ,\ea
where $\tau=\hbar/\left(mc^2\right)$ is a characteristic time, and dots denote derivatives with respect to $t$. Here we actually resort to the mean field
description, similar to that of the Bose-Einstein condensate in \cite{FOOP}. Namely, instead of the quantum potential $Q_1$ of a single particle we
introduce the collective quantum energy density $\varepsilon_q$. Under the assumption of noninteracting particles, this is a well-grounded approach
allowing, in particular, to construct Bohmian hydrodynamics for a perfect fluid \cite{FOOP}.

The corresponding quantum pressure $p_q$ can be found from the first law of thermodynamics written down in the form
$d\left(\varepsilon_qa^3\right)+p_qd\left(a^3\right)=0$:
\ba{6} p_q=\frac{1}{4}\tau^2\varepsilon_{0}\left(\frac{a_0}{a}\right)^3a\frac{d}{da}\left(\frac{\ddot a}{a}+\frac{1}{2}\frac{\dot a^2}{a^2}\right)
=\frac{1}{4}\tau^2\varepsilon_{0}\left(\frac{a_0}{a}\right)^3\frac{a}{\dot a}\left(\frac{\dddot a}{a}-\frac{\dot a^3}{a^3}\right)\, .\ea

It is worth mentioning that the expression \rf{4} represents the direct generalization of the nonrelativistic one-particle quantum potential (described, e.g.,
by the formula (15) in \cite{FOOP}) to the relativistic case: the density $n({\bf r},t)$ is replaced by the relativistic-invariant energy density
$\varepsilon_c$, and the corresponding relativistic one-particle quantum potential \rf{4} is characterized by the correct nonrelativistic limit
$Q_1\rightarrow-[\hbar^2/(2m)]\triangle\sqrt{n}/\sqrt{n}$ when $c\rightarrow+\infty$, $\varepsilon_c\rightarrow mc^2n({\bf r},t)$ (here, of course, the flat
spacetime case is meant).

The Einstein equations give
\ba{7} \frac{3\dot a^2}{c^2a^2}=\frac{8\pi G_N}{c^4}\left(\varepsilon_c+\varepsilon_q\right)=\frac{8\pi
G_N}{c^4}\varepsilon_{0}\left(\frac{a_0}{a}\right)^3\left[1-\frac{3}{4}\tau^2\left(\frac{\ddot a}{a}+\frac{1}{2}\frac{\dot a^2}{a^2}\right)\right]\, ,\ea
where $G_N$ is Newtonian gravitational constant. It immediately follows from \rf{7} that the Hubble parameter squared
\ba{8} H^2&=&\left(\frac{\dot a}{a}\right)^2=\frac{1}{T^2(a)}\left[1+C\exp\left(-\frac{8T^2(a)}{9\tau^2}\right)\right]\, ,\nn\\
T(a)&=&T_0\left(\frac{a}{a_0}\right)^{3/2}=\left(\frac{3c^2}{8\pi G_N\varepsilon_0}\right)^{1/2}\left(\frac{a}{a_0}\right)^{3/2}\, ,\ea
where $T_0=\sqrt{3c^2/(8\pi G_N\varepsilon_0)}$ is one more characteristic time, and $C$ represents some integration constant. It should be noted here that in
the limit $a\rightarrow+\infty$ we get $H^2\rightarrow T^{-2}(a)=8\pi G_N\varepsilon_c(a)/\left(3c^2\right)$, in complete agreement with the corresponding
cosmological model disregarding quantum effects (see, e.g., \cite{Landau,Gorbunov,Rubakov}). Thus, the classical behavior is reproduced asymptotically as it
should be.

Being interested exclusively in the inflationary stage of the Universe evolution, at first let us naively impose a natural initial condition: $H\rightarrow
\lambda=\mathrm{const}>0$ when $a\rightarrow 0$ (the Big Bang moment). Then $C=-1$ and, consequently,
\be{9} H^2=\left(\frac{\dot a}{a}\right)^2=\frac{1}{T^2(a)}\left[1-\exp\left(-\frac{8T^2(a)}{9\tau^2}\right)\right]\, .\ee

For sufficiently small values of $a$ we obtain from \rf{9}
\be{10} H^2\approx \frac{1}{T^2(a)}\left(1-1+\frac{8T^2(a)}{9\tau^2}\right)=\frac{8}{9\tau^2}=\lambda^2\, ,\ee
whence
\be{11} \lambda=\frac{2\sqrt{2}}{3\tau}=\frac{2\sqrt{2}}{3}\frac{mc^2}{\hbar}\, .\ee

It also immediately follows from \rf{9} that
\ba{12} \frac{\ddot a}{a}=\frac{4}{3\tau^2}\exp\left(-\frac{8T^2(a)}{9\tau^2}\right)
-\frac{1}{2T^2(a)}\left[1-\exp\left(-\frac{8T^2(a)}{9\tau^2}\right)\right]\, .\ea

For sufficiently small values of $a$ we get from \rf{12} that $\ddot a/a\approx \lambda^2$. Thus, at the Big Bang moment (we can choose the value $t=0$ for it
without loss of generality), when $a=0$, the ratios $\dot a/a$ and $\ddot a/a$ are both positive and finite. The first fact gives rise to inflation while the
second one ensures smoothness of the scale factor and its derivative even at $t=0$. On the contrary, it is known that in the framework of the corresponding
cosmological model disregarding quantum effects (see, e.g., \cite{Landau,Gorbunov,Rubakov}) $a(t)\sim t^{2/3}$ and, consequently, the derivatives $\dot
a(t)\sim t^{-1/3}$, $\ddot a(t)\sim-t^{-4/3}$ and the ratios $\dot a/a\sim t^{-1}$, $\ddot a/a\sim -t^{-2}$ have singularities at $t=0$. Besides, obviously,
there is no inflation in this case since the expansion of the Universe is decelerating. Taking into account quantum effects eliminates these disadvantages in
elegant manner.

However, from the purely mathematical point of view, the aforesaid naive choice of initial conditions will give $a(t)=0$ forever, because simultaneously
$a(0)=0$, $\dot a(0)=0$ (and $\ddot a(0)=0$). In this connection, as usual, let us impose another initial condition
\be{13} a(t_{\mathrm{Pl}})=a_0,\quad t_{\mathrm{Pl}}=\sqrt{\frac{\hbar G_N}{c^5}}\approx5.391\times10^{-44}\, \mathrm{s}\, ,\ee
which is prevalent and reasonable from the physical point of view \cite{Gorbunov,Rubakov}. Consequently, $a_0=a_{\mathrm{Pl}}$,
\be{14} \varepsilon_0=\varepsilon_{\mathrm{Pl}}=\frac{c^7}{\hbar G_N^2}\approx4.633\times10^{114}\, \frac{\mathrm{erg}}{\mathrm{cm}^3}\, .\ee

It should be mentioned that there is no sense to apply the derived equation \rf{9} for $t<t_{\mathrm{Pl}}$ (Planck epoch). It is applicable only for
$t\geqslant t_{\mathrm{Pl}}$, and we confine ourselves to this case.

Concluding this section, we make an extremely important generalization to the multicomponent case by redefining $\varepsilon_0$ and $\tau$ (as well as $m$) as
follows:
\ba{15} \varepsilon_0=\sum\limits_{i}\varepsilon_{0i},\quad \tau^2=\frac{1}{\varepsilon_0}\sum\limits_{i}\varepsilon_{0i}\tau_i^2,\quad
\frac{1}{m^2}=\frac{1}{\varepsilon_0}\sum\limits_{i}\varepsilon_{0i}\frac{1}{m_i^2}\, ,\ea
where $\tau_i=\hbar/\left(m_ic^2\right)$. In other words, we identify $\varepsilon_0$ with the total energy density of the nonrelativistic particle soup and
$\tau$ (as well as $m$) with its averaged parameters. This generalization is crucial at least for two main reasons. First, the mass generators (Higgs bosons)
must be evidently present in the mixture, but they do not necessarily play the leading or even significant role in the proposed inflation mechanism, perhaps,
letting other coexisting particles have it. Second, if the theory requires a certain value of $m$ which does not correspond to any known particle, there is a
good chance to obtain this required value by mixing different particles with known masses.

Let us briefly illustrate the situation by considering the two-component system of particles. It will be characterized by the mass $m$ defined by the equation
\be{16} \frac{1}{m^2}=\frac{\eta_1}{m_1^2}+\frac{\eta_2}{m_2^2},\quad \eta_1=\frac{\varepsilon_{01}}{\varepsilon_{01}+\varepsilon_{02}},\quad
\eta_2=\frac{\varepsilon_{02}}{\varepsilon_{01}+\varepsilon_{02}}\, ,\ee
where $\eta_1$ and $\eta_2$ are the corresponding energy fractions, $\eta_1+\eta_2=1$. Under which conditions the contribution of particles of the second kind
may be neglected here? Obviously, the answer is provided by the strong inequalities
\be{17} \frac{\eta_1}{m_1^2}\gg \frac{\eta_2}{m_2^2},\quad \eta_1\gg \eta_2\left(\frac{m_1}{m_2}\right)^2\, .\ee

For example, if $m_1\approx 10^{-5}\, \mathrm{eV}$ (according to \cite{RusovPRD,Gorbunov,Rubakov}, this may be the axion mass) while $m_2\approx 125\,
\mathrm{GeV}$ (this value is associated with the Higgs boson mass), then particles of the first kind play the leading role for inflation if
$\eta_1\gg\eta_2\times6.4\times10^{-33}$, and this strong inequality may hold true even when their energy fraction $\eta_1$ is really negligible in comparison
with $\eta_2$. Thus, the major part in the proposed inflation scenario belongs to the lightest particles in the soup.

Evidently, the assumption that the considered particles are nonrelativistic does not contradict the general notion of the ''hot'' early Universe. Really, if
particles of some sort are initially ''cold'' and interact weakly enough with particles of all other sorts, then they will remain nonrelativistic. At the same
time the average temperature of the whole soup can be high.

%\vspace{0.3cm}

\section*{Duration of inflation and number of e-folds}

In order to define duration of inflation, let us answer the following important question: at which moment does the expansion acceleration equal zero? This
moment can be found from the equation
\be{18} 3\mu_e\exp\left(-\mu_e\right)-1+\exp\left(-\mu_e\right)=0,\quad \mu_e=\frac{8T^2(a_e)}{9\tau^2}\, ,\ee
where $a_e=a(t_e)$. The numerical solution is $\mu_e\approx1.904$. Obviously, the less is $\tau$, the less is $a_e$ (as well as the time $t_e$ itself) required
for reaching this value of $\mu$. The equation
\ba{19} \frac{T^2(a_e)}{\tau^2}=\frac{3c^2}{8\pi G_N\varepsilon_{\mathrm{Pl}}\tau^2}\left(\frac{a_e}{a_{\mathrm{Pl}}}\right)^3=
\frac{3}{8\pi}\left(\frac{m}{m_{\mathrm{Pl}}}\right)^2 \left(\frac{a_e}{a_{\mathrm{Pl}}}\right)^3\approx2.142 \ea
defines the end of the inflation stage. Here
\be{20} m_{\mathrm{Pl}}=\sqrt{\frac{\hbar c}{G_N}}\approx2.177\times10^{-5}\, \mathrm{g}=1.221\times 10^{28}\, \mathrm{eV}\, .\ee

It follows from \rf{19} that the number of e-folds is given by a very simple and elegant expression:
\be{21} N_e=\ln\frac{a_e}{a_{\mathrm{Pl}}}\approx 0.9624+\frac{2}{3}\ln\frac{m_{\mathrm{Pl}}}{m}\, .\ee

The accepted range $50\lesssim N_e\lesssim60$ \cite{PLANCK} corresponds to the mass range $4.238\times10^{-11}\, \mathrm{eV}\lesssim m\lesssim 1.385\times
10^{-4}\, \mathrm{eV}$. It is interesting that the axion mass $m_a\approx 10^{-5}\, \mathrm{eV}$ \cite{RusovPRD,Gorbunov,Rubakov} lies within this range and
gives $N_e\approx 52$. It means that if the axion is a driving force of inflation, then its maximum energy fraction $\eta_a^{\mathrm{max}}$ may tend to a
unity. At the same time, the minimum axion energy fraction $\eta_a^{\mathrm{min}}$ may equal approximately $0.005$ (i.e., only $0.5\,\%$), corresponding to
$N_e\approx50$. So it is enough to have only $0.5\,\%$ (with respect to the energy density) of axions in the mixture with $99.5\,\%$ of other heavier particles
for ensuring successful inflation.

It is also interesting that ultralight particles (considered, e.g., in \cite{LeeLim,Lukewarm,Hu} as other dark matter candidates) with the mass $m_{ul}$ of the
order $10^{-(22\div23)}\, \mathrm{eV}$ may serve as a suitable driving force of inflation with negligible energy fractions $5.21\times 10^{-(37\div39)}\lesssim
\eta_{ul}\lesssim 5.57\times10^{-(24\div26)}$.

Since $\tau/t_{\mathrm{Pl}}=m_{\mathrm{Pl}}/m$, we also easily get the $\tau$-range $4.751\times10^{-12}\,\mathrm{s}\lesssim \tau \lesssim
1.553\times10^{-5}\,\mathrm{s}$. At the same time, for the $t_e$-range we have $2.528\times10^{-10}\,\mathrm{s}\lesssim
t_e\lesssim9.912\times10^{-4}\,\mathrm{s}$.

Introducing the convenient dimensionless quantities
\be{22} \mu(a)=\xi^3(a)=\frac{8T^2(a)}{9\tau^2}=\frac{1}{3\pi}\left(\frac{m}{m_{\mathrm{Pl}}}\right)^2 \left(\frac{a}{a_{\mathrm{Pl}}}\right)^3\, ,\ee
from \rf{9} and \rf{12} we obtain respectively
\be{23} \frac{H}{\lambda}=\left[\frac{1-\exp\left(-\xi^3\right)}{\xi^3}\right]^{1/2}\, ,\ee
\be{24} \frac{\ddot a}{\lambda^2a}=\frac{3}{2}\exp\left(-\xi^3\right)-\frac{1}{2\xi^3}\left[1-\exp\left(-\xi^3\right)\right]\, .\ee

Both these functions are shown in Fig. 1 (blue and orange curves respectively). Red and purple curves correspond to the classical case without quantum effects
when all terms containing exponential functions are dropped and $H=\lambda\xi^{-3/2}$, $\ddot a/a=-\left(\lambda^2/2\right)\xi^{-3}$. Finally, the green
vertical line $\xi=\xi_e=(\mu_e)^{1/3}$ signifies the end of inflation. So there are inflationary expansion (with the positive acceleration) for $\xi<\xi_e$
and traditional expansion (with the negative acceleration) for $\xi>\xi_e$, as it certainly should be.

\begin{figure}[htbp]
\begin{center}\includegraphics[width=3.2in,height=2.4in]{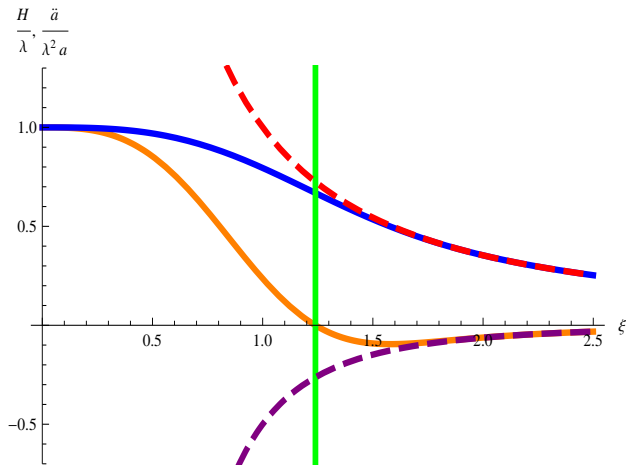}\end{center}
\caption{Ratios $H/\lambda$ \rf{23} and $\ddot a/\left(\lambda^2a\right)$ \rf{24} as functions of $\xi$.}
\end{figure}

%\vspace{0.3cm}

\section*{Summary}

In this paper we have constructed a simple cosmological model of the Universe filled with a soup of nonrelativistic particles. The contribution of their
quantum potential to the corresponding energy-momentum tensor (see \rf{5} for the quantum energy density $\varepsilon_q$) has led to the scale factor $a(t)$
(see \rf{9} for the Hubble parameter squared $H^2$ and \rf{12} for the ratio $\ddot a/a$), demonstrating the inflationary behavior $\dot a\sim a$ immediately
after Planck epoch and then approaching asymptotically the classical (i.e., non-quantum) limit $\dot a\sim a^{-1/2}$.

Our inflationary theory gives the reasonable number of e-folds $50\lesssim N_e\lesssim60$ for a specified range of the effective particle mass
$4.238\times10^{-11}\, \mathrm{eV}\lesssim m\lesssim 1.385\times 10^{-4}\, \mathrm{eV}$. We have shown that the lightest particles in the soup most likely
play a crucial role in this scenario (see \rf{15}-\rf{17} and the related text). Axions and ultralight particles represent the illustrative examples, for
which we have imposed constraints on the corresponding energy fractions.

Thus, we propose a promising mechanism of successful inflation. Of course, cosmological problems of the post-inflationary stage as well as an explanation
for the origin of primordial fluctuations and necessary predictions of their statistical properties, which are currently tested by observations of the
cosmic microwave background anisotropy, lie beyond the scope of this short paper and require separate additional research. Our main aim was to change the
angle of view on cosmology of the earliest Universe. Drawing a conclusion, we claim that the quantum potential of light particles may serve as the main
reason for its inflationary evolution.

%\vspace{0.3cm}

\section*{Acknowledgements}

The work of M. Eingorn was supported by NSF CREST award HRD-1345219 and NASA grant NNX09AV07A.

%\vspace{0.3cm}

\end{document}